\documentclass[useAMS,usenatbib]{mn2e}

\usepackage{graphicx,times}

\title[High redshift galaxies with TP-AGB stars: comparison with observations]
      {Hierarchical models of high redshift galaxies with thermally pulsing asymptotic giant branch stars: comparison with observations}

\author[C. Tonini et al.]
{Chiara Tonini$^{1}$
\thanks{E-mail:chiara.tonini@port.ac.uk},
Claudia Maraston$^{1}$,
Daniel Thomas$^{1}$,
Julien Devriendt$^{2}$
\newauthor
and Joseph Silk$^{2}$ \\
$^{1}$Institue of Cosmology and Gravitation, University of Portsmouth, PO1 3FX Portsmouth, UK\\
$^{2}$University of Oxford, OX1 3PU Oxford, UK\\
}

\begin{document}



\maketitle

\begin{abstract}

In a recent paper we presented the first semi-analytic model of galaxy formation in which 
the Thermally-Pulsing Asymptotic Giant Branch
phase of stellar evolution has been fully implemented. Here we address the comparison with 
observations, and show how the TP-AGB recipe
affects the performance of the model in reproducing the colours and near-IR luminosities
of high-redshift galaxies. 
We find that the semi-analytic model with the TP-AGB better matches 
the colour-magnitude and colour-colour
relations at $z \sim 2$, both for nearly-passive and for star-forming galaxies.
The model with TP-AGB produces star-forming galaxies with red V-K colours, 
thus revising the unique interpretation
of high-redshift red objects as 'red \& dead'. We also show that without the 
TP-AGB the semi-analytic model 
fails at reproducing the observed colours, a situation that cannot be corrected by
dust reddening.
We also explore the effect of nebular emission on the predicted colour-magnitude relation
of star-forming galaxies,
to conclude that it does not play a significant role in reddening their colours, 
at least in the range of star-formation rates covered by the model.
Finally, the rest-frame K-band luminosity function at $z \sim 2.5$ is more luminous by almost
1 magnitude. This indicates that the AGN feedback recipe that is adopted to regulate
the high-mass end of the luminosity function should be sophisticated to take 
the effect of the stellar populations into account at high redshifts.

\end{abstract}


\section{Introduction}

Hierarchical clustering is the favoured scenario to describe
the formation and evolution of matter structures in the universe 
(White \& Rees, 1978), and semi-analytic models of galaxy formation 
proved themselves to be a powerful tool of investigation since the first 
formulation (White \& Frenk, 1991). Over the years, many
such models have been developed (see for instance Balland et al. 2003, 
Baugh et al. 2005, Bower et al. 2006, 
Cattaneo et al. 2008, Cole et al. 2000, De Lucia et al. 2004, Hatton et al. 2003, 
Kauffmann et al. 1993, Menci et al. 2006, Monaco et al. 2007, Somerville et al. 2008). 
The successes and failures of these models are strictly linked to those of
the hierarchical scenario itself, ultimately depending on the mechanisms of
mass accretion of objects as a function of time. The large-scale
structure and the integrated properties of the galaxy population (such as the total stellar
mass density for instance) are well reproduced. 
The detailed evolution of galaxies however presents several puzzling aspects, 
such as e.g. the size of the disks of spirals (Burkert 2009), or the $\alpha$-enhancement 
and the ages of the stellar populations in massive ellipticals (Thomas 1999, 
Thomas et al. 1999, Cimatti et al. 2004, Nagashima et al. 2005, 
Thomas et al. 2005, Pipino et al. 2009, Kormendy et al. 2009). 
Moreover, global properties of the galaxy population such as the
evolution of the stellar mass function (Cole et al. 2001, Bell et al. 2003, Bundy et al 2006-2009, 
Pozzetti et al. 2009, Colless et al. 2007) are still not reproduced in the models
(Bundy et al. 2007, Marchesini et al. 2009, Kajisawa et al. 2009, Kodama \& Bower 2003), 
although there is controversy on this point (Drory et al. 2004, Benson et al. 2007).
It is debated if any of these problems 
can possibly be resolved with enhanced resolution
in the simulations and more sophisticated recipes in the models. 

One of the most problematic issues for the models is to reproduce
the abundance of high redshift luminous galaxies 
(e.g., Conselice et al. 2007, Cimatti et al 2004, van Dokkum et al. 2004, 2006). 
This difficulty is partly due to a mis-interpretation
of the nature of these objects.

The high-luminosity end of the galaxy population up to redshift $z \sim 2.5$ 
consists in fact of objects that look like the early-type galaxies in the local 
universe, \textit{i.e.} they are characterized by very red colours in the optical
and near-IR, and high near-IR luminosities (Mancini et al. 2009, Cimatti et al 2004, 
McCarthy et al. 2004, Daddi et al. 2005, Saracco et al. 2005, Kriek et al. 2006).
Local ellipticals showing the same photometry are old 
(with stellar populations older than $\sim1$ Gyr), 
passively evolving, and with stellar masses $M_{star} > 10^{11} \ M_{\odot}$.
Moreover, newer observations
are building the case for the presence of extremely red and IR-luminous objects at even higher redshifts 
(Rodighiero et al. 2007, Mancini et al. 2009, Fontana et al. 2009).

The problem posed by the presence of these high redshift ($z>2$) red and luminous galaxies 
stems from the consensus that they are massive objects evolving
passively, the so-called 'red \& dead' galaxies.  
With the stellar population models currently used in the 
semi-analytic models in the literature (for the most part 
Bruzual \& Charlot 2003, hereafter BC03), 
the only way to explain the high near-IR luminosities of high-redshift
galaxies is to advocate very high galaxy masses and very old ages
of the stellar populations. But these are not achieved in the actual 
model realizations (except at low redshifts), because the 
hierarchical mass assembly has an intrinsic difficulty in putting 
together massive and old objects at early epochs. In fact, the hierarchical scenario
predicts a steady decline of the abundance 
of massive galaxies with increasing redshift (van Dokkum et al. 2004).

In Tonini et al. (2009) we showed that the predictions of colours and 
luminosities of galaxies at high redshift in a semi-analytic model
are greatly affected by the recipes in use for the stellar populations, 
expecially the inclusion of the Thermally-Pulsing Asymptotic Giant Branch (TP-AGB).
As shown in M05, in stellar populations of intermediate age ($\leq 0.2 - 2$ Gyr)
the TP-AGB phase dominates the near-IR luminosity,
with a contribution up to $80 \%$ in the rest-frame K band, and contributes to up to
$40 \%$ of the bolometric luminosity (M05).
High-redshift galaxies, in which the mean age of the stellar populations
is in that range, are expected to be dominated by the TP-AGB emission in
the near-IR. This has been recently confirmed by SED-fitting of observations 
made with the Spitzer Space Telescope (Maraston et al. 2006, Cimatti et al. 2008).  
In Tonini et al. (2009) we included a complete treatment of the TP-AGB phase in the semi-analytic
model GalICS (Hatton et al. 2003), by implementing the M05 stellar
population models into the code, and showed that the rest-frame 
$V-K$ colours at high redshift get redder by more than 1 magnitude. 
Relatedly, the K-band mass-to-light ratio is 
shifted towards luminosities 1 magnitude higher for a given galaxy mass. 
Notably, actively evolving, star-forming high-redshift galaxies are predicted to
have $V-K$ colours and near-IR luminosities similar to those of local, 
passively evolving massive systems (Tonini et al. 2009).

Once the stellar emission is correctly modeled with an exhaustive treatment
of all the significant phases of stellar evolution, a more accurate comparison
between the semi-analytic model predictions and the data is possible. 
In particular, the 
performance of the semi-analytic model in reproducing the observed colours and luminosities
in the near-IR becomes meaningful to test the hierarchical mass assembly at different 
redshifts.

In the literature the comparison between galaxy formation models and data is typically 
done by obtaining physical properties for the real objects through application of stellar 
population models to data. However, this approach carries several degeneracies, including 
the adopted population synthesis model, the recipe for star formation history, the choice 
of metallicity, etc. When a realistic errorbar including all these variables ia attached 
to the observationally derived quantity, such as in Marchesini et al. 2009 (and see also 
Conroy et al. 2009 for a discussion), the results of such comparisons may not be clear-cut.

In this paper we adopt a different philosophy for the 
comparison between model and data. Instead of using processed data in the
rest-frame system, we consider raw, unprocessed, apparent magnitudes straight
out of the catalogues. We then produce mock catalogues out of the simulation, 
so that the output spectra of the model galaxies are redshifted in the observer's frame. 
The model apparent magnitudes and colours can then be directly compared with 
the observational data. This comparison yields direct information about the 
physical quantities in the model in use.  

This procedure is straightforward and does not add substantial degeneracy that can 
jeopardize the comparison. 
A degeneracy that clearly remain is how dust reddening affects the intrinsic stellar 
emission, as recently pointed out by Guo \& White (2008) and Conroy et al. (2009). 
However, we shall show that considering the intrinsic star formation rates in the 
model and using data mapping the rest-frame near-IR, such an uncertainty plays actually a minor role.

The structure of the paper is as follows.
In Section 2 we briefly introduce the new semi-analytic model GalICS with the TP-AGB implementation
through the M05 models (as from Tonini et al. 2009). In Section 3 we 
describe the data samples used for our analysis. In Section 4 we compare
the colour-magnitude and colour-colour relations predicted by the model 
against samples of $z \sim 2$ galaxies. In Section 4 we compare the model
rest-frame K-band luminosity function in the M05 and Pegase cases with the predictions
by other semi-analytic models.
In Section 5 we discuss our results.    

\section{The semi-analytic model of galaxy formation} 

We produce the model galaxies through the hybrid semi-analytic model
GalICS (Hatton et al. 2003), and we defer the reader to its original
paper for details on the dark matter N-body simulation and the 
implementation of the baryonic physics. In brief, the model builds up the galaxies
hierarchically, and evolves the metallicity consistently with the 
cooling and star formation history (with the new implementation by Pipino et
al. 2009). Feedback recipes for supernovae-driven
winds and AGN activity are implemented in the code (the lattest with the improved
version of Cattaneo et al. 2006). Merger-driven
morphology evolution and satellite stripping and disruption are taken into account.

The semi-analytic model was originally supplied with two 
different sets of input stellar population models, namely the PEGASE (Fioc \& Rocca-Volmerange, 1997)
and the Stardust (Hatton et al. 2003, Cattaneo et al. 2008). The Stardust model is  
too rudimentary for the scope of this work, so we discarded it. 
In Tonini et al. (2009) for the first time the M05 models were implemented into GalICS, 
and the predictions of the models were compared with those obtained with PEGASE.
The same sets of SSP models are utilized in the present paper. 
For the purpose of this work, the most significative difference between PEGASE and M05
is that the M05 includes higher energetics for the TP-AGB phase. The PEGASE
models in this respect produce results comparable to the more commonly used 
BC03 models (see M05), with a TP-AGB recipe with much lower energetics 
than the M05 (see M05, Fig. 18).
For both sets of SSPs (Single Stellar Populations), the adopted IMF is Salpeter, and 
the metallicity range is $0.001 < Z < 0.04$ (where $Z_{\odot}=0.02$).

In the current implementation, dust extinction is taken into account. 
The model spectra are reddened according to the ongoning star formation.
We adopt a Calzetti extinction curve and a colour-excess $E(B-V)$
proportional to the star formation rate for each single galaxy, 
parameterized as $E(B-V) = 0.33 \cdot (Log(SFR)-2) + 1/3$. 
This choice is supported by data analysis in the literature, in general for 
samples of star-forming galaxies at redshifts around $z \sim 2$ (Daddi et al. 2007, 
Maraston et al., in prep.), 
and in particular it agrees with the SED fitting of the galaxies in the data samples 
used in this paper, which are introduced in the next Section.
We do not randomize the inclination of disk galaxies (which reduces the dust effect
in face-on objects) but we redden the spectrum 
by the total amount of extinction calculated for each galaxy,
therefore considering the maximal reddening for each object. 
As it will be clear in the next Sections,
this choice proves to be instructive
in the comparison between GalICS runs with M05 and PEGASE input SSPs.    
This reddening recipe is more sophisticated than simple screen-models used so far, 
and follows the spirit of important developments in the field
(see for example Ferrara et al. 1999, Guo $\&$ White 2008). The results presented here have 
been tested against different extinction curves (Large Magellanic Cloud, Milky Way and 
Small Magellanic Cloud types), with the 
conclusion that the main factor affecting the dust contribution is $E(B-V)$.  

The comparison with data performed in this work requires the distinction between
actively star-forming galaxies and nearly-passive galaxies (passively evolving or
with small residual star formation). For this purpose, we split our model
galaxies according to their istantaneous star formation rate, with the 
criterium that objects with SFR $\le 3 \ M_{\odot}/yr$ are considered nearly-passive, 
and galaxies with SFR $> 10 \ M_{\odot}/yr$ are actively star-forming. 
This value was chosen according to the star-forming
objects selected in Daddi et al. 2007 and in Maraston et al. 
(in preparation), where SFR $>10 \ M_{\odot}/yr$ are robustly 
determined from far-IR and the UV-slope method.
 
We build mock catalogues from the simulation, by redshifting the rest-frame spectra 
at each timestep, to produce observer-frame 
luminosities and colours, to compare directly with the data. We set the
model magnitudes to mimic the data catalogues in use, by 
filtering the spectra with the same broadband filters used in the observations. 
The broadband magnitudes thus obtained are further scattered with gaussian errors comparable to the 
observational errors of our data samples (on average $\sigma = 0.1 mag$ at $z \sim 2$).

\section{Data selection}

We want to compare the predicted colours and near-IR luminosity of our 
semi-analytic model against data of high-redshift galaxies, 
to test whether our improved stellar populations implementation
allows to better match the properties of the red galaxy population.
Given the nature and origin of the TP-AGB light, we focus our analysis
on datasets for which the IR photometry is available. In fact, the optical to near-IR rest-frame
emission will get redshifted into the IRAC bands for redshifts $z \geq 2$.
We explicitly looked for samples of excellent photometry quality, 
in order to minimize any source of uncertainty. 

The first sample is from Maraston et al. (2006). It consists of 7 galaxies
selected in the optical by Daddi et al. (2005), with photometry extended 
to cover the rest-frame near-IR. These galaxies were selected 
 from the Hubble Ultra Deep Field through the $BzK$ technique
introduced by Daddi et al. (2004), and have spectroscopic redshifts 
between $1.4 \leq z \leq  2.7$ (see Maraston et al. 2006 for details).
They show early-type morphology, and SED fitting (Maraston et al. 2006) indicates
an average age between $0.2 < \tau < 2$ Gyr and masses $\sim 10^{11} \ M_{\odot}$, 
a modest dust attenuation and a negligible amount of OB stars, so that these 
objects qualify as nearly passive. In Daddi et al. (2005) they were interpreted
as progenitors of present-time massive ellipticals. 
The second sample is taken from the 
GOODS-S (Great Observatories Origins Deep Survey -South) catalogue from Daddi et al. (2007),
and consists of 95 galaxies with high-quality photometry and spectroscopic 
redshifts between $1.7 \leq z \leq  2.3$. The $BzK$ method identifies them as
star-forming, which has been confirmed by SED fitting (Maraston et al., in prep.). 

\section{Colour-magnitude and colour-colour relations at $z \sim 2$}

In this Section we compare the observed broadband magnitudes and colours of the data
samples with the predictions of the model. We translate the model results in the 
observer-frame at each redshift, so that all magnitudes presented here are apparent.
The magnitudes are calculated in the AB system.

\begin{figure*}
\includegraphics[scale=0.9]{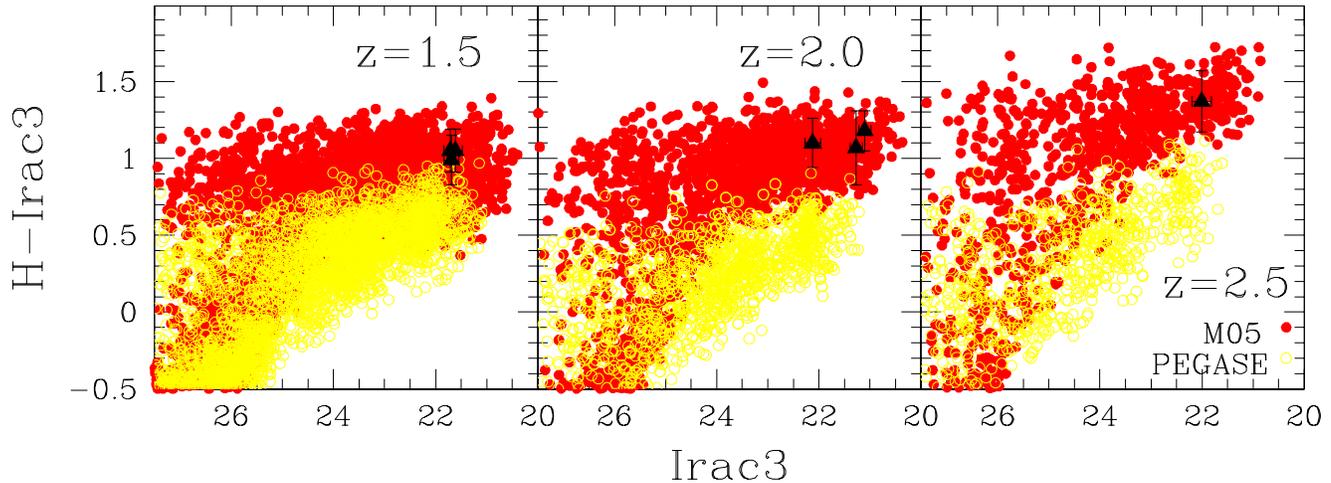}
\caption{The observed-frame colour-magnitude relation Irac3 $vs$ H-Irac3 
corresponding to rest-frame K $vs$ V-K at redshifts $z=1.5, 2, 2.5$ 
(\textit{from left to right}), for nearly-passive galaxies in the M05 runs
(\textit{red filled dots}) and the PEGASE runs (\textit{yellow empty dots}), compared with 
data of nearly passively-evolving galaxies from Maraston et al. (2006) (\textit{black triangles 
with errorbars}). The dust reddening adopted in the models is a LMC-type law
with colour-excess $E(B-V)=0.2$ at $z=1.5$, and a Calzetti law with $E(B-V) \propto SFR$
at $z=2,\ 2.5$.}
\label{7gals}
\end{figure*}

\subsection{Nearly passive galaxies}

Interesting candidates for studying the effects of the TP-AGB in the semi-analytic model
are galaxies dominated by intermediate-age stellar populations. 
In such objects most of the stellar component is active in the TP-AGB 
phase, thus maximizing its effect, 
while star formation is subdominant, which reduces the complication of dust. 
In Fig.~(\ref{7gals}) we compare the model 
observed-frame colour-magnitude relation Irac3 $vs$ H-Irac3,
corresponding to rest-frame K $vs$ V-K, for the nearly-passive galaxies, with data of the 7 galaxies 
at redshifts $z=1.5, 2, 2.5$ singled out by Maraston et al. (2006). 
The model galaxies are plotted as \textit{red filled dots} for the TP-AGB run (M05) and 
\textit{yellow empty dots} for the PEGASE run, while the data are represented by
\textit{black triangles with errorbars}.

The comparison between model and observations show that the M05 run with the TP-AGB
perfectly matches the data, in the whole redshift range. With PEGASE on the
other hand, the model galaxies feature much bluer colours, with an offset of about 
0.5 mags on average between the two runs. In the central panel ($z=2$) it is
also evident that the PEGASE run fails to reproduce the Irac3 (rest-frame K) 
luminosity of these objects, while the M05 run produces luminosities up to 1 mag higher
and easily accomodates the observed ones.

\begin{figure*}
\includegraphics[scale=0.9]{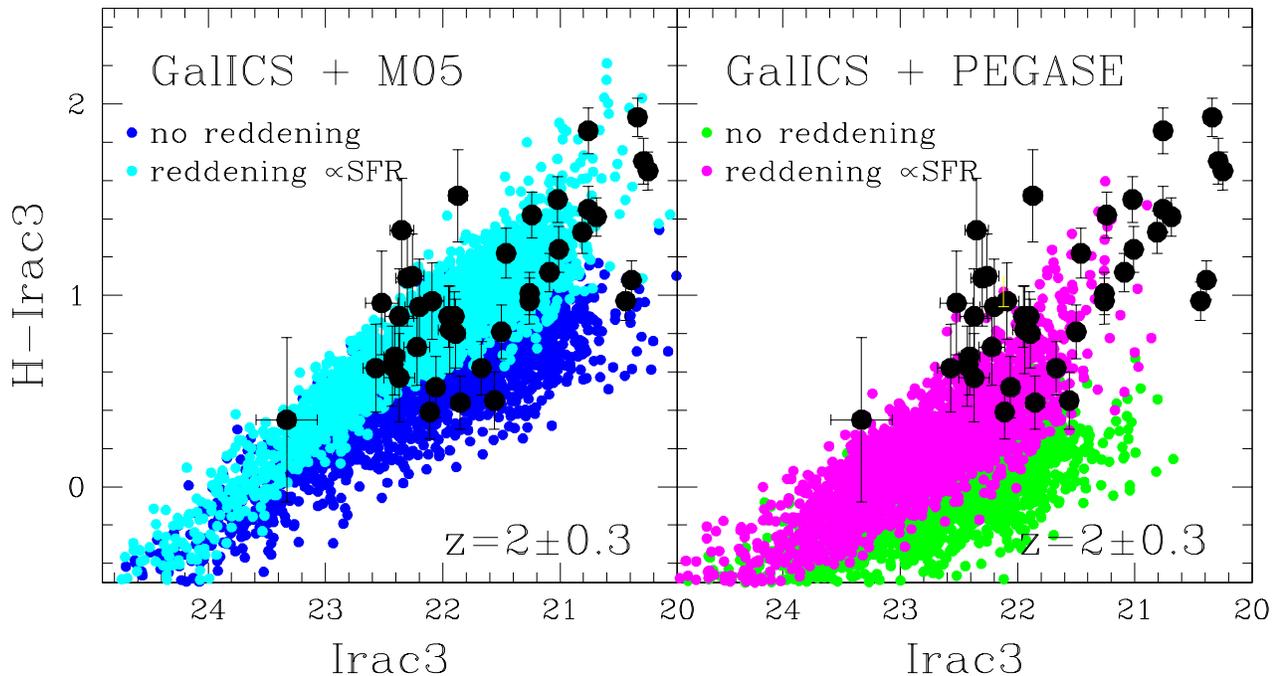}
\caption{The theoretical colour-magnitude relation Irac3 $vs$ H-Irac3 at $z=2$, corresponding
to the rest-frame K $vs$ V-K, for the star-forming galaxies in the M05 run (\textit{left panel}) and
the PEGASE run (\textit{right panel}), compared with data. 
The \textit{black dots} with errorbars are data
points from the GOODS-S catalogue (Daddi et al. 2007). 
In both panels, the results are shown for runs with reddening (a Calzetti-type
extinction with colour-excess proportional to the SFR; \textit{cyan and magenta dots}) 
and without reddening (\textit{blue and green dots}).}
\label{colmag}
\end{figure*}

These galaxies show early-type morphology and lack any significant
emission from young stars. However, as expected at these high redshifts, 
their star formation is not strictly zero\footnote{This 
actually suits the semi-analytic model, since the hierarchical nature of 
the simulation causes satellites to continually infall into bigger objects, 
triggering spurious star formation. Residual star formation is also caused by 
the cooling of hot halo gas onto the central objects, but while at these high redshifts
this is acceptable, it becomes a problem of the model at lower redshifts 
(the so-called cooling catastrophe).}, hence the label 'nearly passive'.
Therefore, although dust reddening in these objects is modest in general, it is not negligible.
The degree of residual star formation in our sample is variable, and in particular
the galaxies at $z=1.5$ show a little bit more activity. In fact, they are SED-fitted  
with a colour-excess due to dust
reddening even slightly higher than predicted by our dust model, so 
for the plot at $z=1.5$
we adopted a dust-screen LMC-type law with $E(B-V)=0.2$ (from Maraston et al. 2006).
Even with a higher reddening however, the PEGASE run is not able to match these data. 

\begin{figure*}
\includegraphics[scale=0.9]{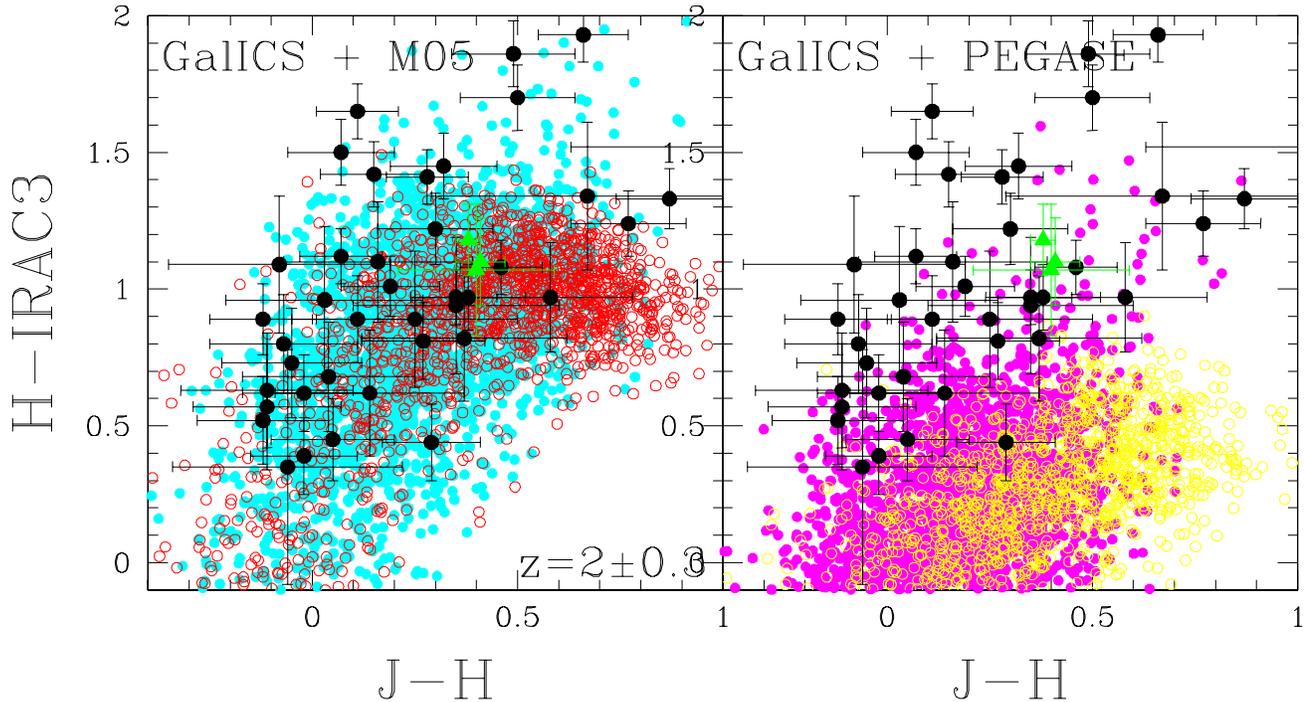}
\caption{The colour-colour relation J-H $vs$ H-Irac3 at $z=2$, corresponding to the
rest frame B-V $vs$ V-K, for the M05 run (\textit{left panel}) and the PEGASE run 
(\textit{right panel}). The \textit{black dots} with errorbars are data
points from the GOODS catalogue (Daddi et al. 2007). The \textit{green triangles} 
with errorbars are data from Maraston et al. (2006) of nearly-passively evolving galaxies. 
For the M05 run, the \textit{red dots} represent nearly-passive galaxies (SFR$<3 \ M_{\odot/yr}$) and 
\textit{cyan dots} represent the star-forming galaxies (SFR$>10 \ M_{\odot/yr}$); in the PEGASE
run, the same holds for \textit{yellow/magenta dots} respectively.}
\label{colcolmiocalzetti}
\end{figure*}

The observed galaxies lie among the most luminous galaxies in the model, as expected.
The slight luminosity overshooting of the simulated galaxies probably stems from the fact that
the simulated volume is much bigger than the survey (at $z=2$ the ratio of the areas
is a factor $\sim 107$). Moreover, the observed galaxies were selected in observed-frame
K band, which maps the $\sim$R at $z=2$, while we are plotting the Irac3 luminosity (rest-frame near-IR). 
The stellar masses at $z \sim 2$ are estimated to be between $10^{10}-10^{11} \ M_{\odot}$,
and are in the range of masses produced by the semi-analytic model at the same
redshift. This match insures a fair comparison between the model stellar luminosities
and the data, by nailing down the mass-to-light ratios and thus
leaving no room for degeneracies in this sense (see Conroy et al. 2009). 
We notice that these galaxies lie close to the top-mass end of the model
distribution; for instance, at $z=2.5$ the observed galaxy lies among the $7 \%$ 
most massive galaxies in the simulation (this amounts to $\sim 500$ objects in the simulated
volume). This shows that the semi-analytic model is performing well in terms of galaxy masses at 
these redshifts, but that there is no margin for reaching the observed colours 
without the TP-AGB. 
We conclude that the TP-AGB appears to be a necessary ingredient of the model
in order to reproduce the colours and near-IR luminosity of these nearly-passive galaxies. 

\subsection{Star-forming galaxies}

Fig.~(\ref{colmag}) shows the same plot as Fig.~(\ref{7gals}) for the model star-forming galaxies,
compared with the sample of star-forming galaxies 
from GOODS, selected in the range $1.7 \leq z \leq 2.3$. The M05 run 
portrayed in the \textit{left panel}, where \textit{cyan dots} represent
the prediction of the semi-analytic model with dust reddening, and the \textit{blue dots}
represent the case without reddening. The \textit{right panel} shows the
PEGASE run, with \textit{magenta dots} for the case with dust reddening, and \textit{green dots} 
for the case without reddening. 

As expected, the galaxies in the M05 run are much redder than in the PEGASE run
(Tonini et al. 2009). They are in excellent agreement with the data.
The M05 run reproduces both the observed amplitude 
of the $H-Irac3$ colour and the slope of the colour-magnitude relation, 
indicating that the SFR across the observed mass range is well represented in the model.
In fact, the model star-formation rates easily cover the range of the ones
derived from this sample of observed galaxies.

The PEGASE run on the other hand is completely off the data, producing much bluer
colours. Moreover, while the Irac3 luminosity range is correctly reproduced  
by the M05, in the PEGASE run the model galaxies are much fainter, and the run 
misses half of the sample in luminosity.
This happens because the TP-AGB increases the emission in Irac3 (rest-frame K) by 1 mag on average.
About $\sim 53.5 \%$ of the galaxies in the GOODS sample have magnitudes 
$Irac3 \leq 22$, and these objects can be reproduced only with the M05 run, while the
run with the PEGASE recipe is far off the mark. On the other hand, only about the most luminous $\sim 10 \%$
of objects is not reproduced by the M05.
  
It is also clear that a higher dust reddening cannot be advocated to make up for the 
absence of the TP-AGB emission and match the observations. 
First of all, dust reddening tilts the colour-magnitude relation upwards
at the high-mass end, but does not increase the Irac3 (rest-frame K) luminosity.
Secondly, the magnitude of its effect on the H-Irac3 colour is well below the 
shift introduced by the TP-AGB. Our reddening recipe is
physically associated with the intrinsic SFR of the model galaxies (a choice sustained
by observations), which at $z \sim 2$ spans a range of values up to a few $10^2 M_{\odot}/yr$, 
corresponding to $E(B-V) \sim 0.3-0.4$ (e.g. Daddi et al. 2007), and yielding values of 
$E(B-V) \sim 0.1-0.2$ for typical SFR produced by GalICS.
Either due to the limited mass resolution of the N-body simulation and/or the recipes
currently employed in GalICS, the hybrid model is not able to produce very massive 
starbursts, so that the maximum model SFR might be
on the low side. The investigation of this issue is beyond the scope of this paper, 
and we shall pursue it further in future work.

Fig.~(\ref{colcolmiocalzetti}) shows the colour-colour relation J-H $vs$ H-Irac3 at $z=2$
(corresponding to rest-frame B-V $vs$ V-K), 
and compares again the M05 run (\textit{left panel}) and the PEGASE run
(\textit{right panel}) with the same set of data. In the panels, 
\textit{red/yellow dots} represent nearly-passive galaxies (SFR$<3 \ M_{\odot/yr}$) and 
\textit{cyan/magenta dots} represent the star-forming galaxies (SFR$>10 \ M_{\odot/yr}$). 

The striking feature highlighted by this plot is that, with the M05 stellar populations, 
the semi-analytic model is now able to produce star-forming galaxies that
are very blue in the optical and very red in the near-IR, and reproduces the data 
very well. The run with the PEGASE recipe on the other hand, while covering the correct 
range in $J-H$, is clearly off the data in the near-IR. Given that the optical 
colours are correctly reproduced, we conclude that a higher amount of reddening
would be unrealistic.  

This plot highlights the ability of GalICS to reproduce the mix of stellar populations
of different ages in these galaxies, by matching the optical blue colours. At the same time, the 
GalICS $+$ M05 run can also match the near-IR colours thanks to the contribution 
of the TP-AGB. Finally, the dust extinction adopted in this implementation lies in a
sensible range of values. Our model reproduces the properties of the observed galaxies at $z \sim 2$
without invoking unrealistic star-formation rates or dust reddening.

Notice the 3 nearly-passive galaxies of Maraston et al. (2006) in the plot 
(\textit{green triangles}). As 
expected, they are near the reddest in optical colours. Most of the sample of star-forming
galaxies is bluer than these objects, but the near-IR colours are comparable. 
Correspondingly in the model, the nearly-passive galaxies are among the reddest 
both in the near-IR and optical colours, but they are not distinguishable from 
active galaxies on the basis of colours alone. 
As already pointed out in Tonini et al. (2009), star formation does not dilute the TP-AGB
emission in the near-IR, so that the near-IR colours alone do not discriminate
between star-forming and passive. In fact, star-forming galaxies can be redder than passive
ones in the near-IR.
This proves that the so-called 'red \& dead' galaxies may not, in fact, be dead at all. 
The same near-IR colours can be achieved in star-forming galaxies, because the 
TP-AGB emission is not offset by the light produced by young stars.

\subsubsection{Effects of removing the TP-AGB from M05 models}

\begin{figure}
\includegraphics[scale=0.4]{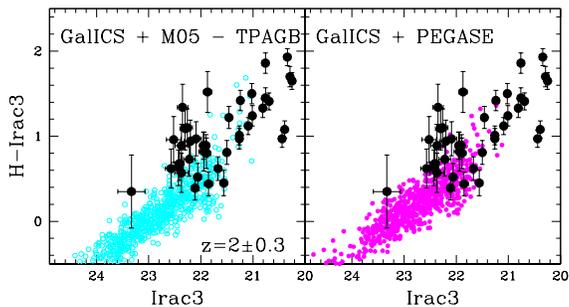}
\caption{Same plot as in Fig.~(\ref{colmag}), but with the comparison between 
the PEGASE run (\textit{right panel}) and a run with Maraston models
without the TP-AGB phase (\textit{left panel}), 
and the rest of the recipe unchanged. In both cases we adopt
the reddening described previously. The (\textit{black dots})
represent the same set of data of Fig.~(\ref{colmag}).}
\label{colmag2}
\end{figure}

Obviously, the M05 and the PEGASE models differ also for other recipes than 
the TP-AGB implementation. For instance, in M05 young stellar populations
are modelled with the Geneva stellar evolutionary tracks, while the Padova tracks
are used in PEGASE. 
In order to rule out that the significant difference in the predictions of colour and 
luminosity shown in Fig.~(\ref{colmag}) is due to any effect other than the TP-AGB, 
 we performed a test run of
GalICS with Maraston (2005) models \textit{without the TP-AGB}, and with the rest of the recipe
unchanged. This quantifies the actual contribution of the TP-AGB in the predictions of the M05 run. 
The result is shown in Fig.~(\ref{colmag2}), where this new run (represented by 
\textit{cyan empty dots, left panel})
is compared with the PEGASE run (\textit{magenta dots, right panel}) 
runs, and the same sets of data. 
The predicted luminosity and colour in the run with Maraston models without TP-AGB is 
strikingly similar to those of the PEGASE run.
This shows that the TP-AGB emission is the main driver
of the success of the model in correctly predicting luminosities and colours of the observed
galaxies. 

\subsubsection{Effects of nebular emission}

It has been pointed out (Leitherer et al. 1999, Zackrisson et al. 2008, Molla' et al. 2009)
that, among the factors that can affect the colours of star-forming galaxies, 
nebular emission can play a significant role. This kind of emission is produced by 
ionizing photons emitted from massive young stars when scattering with the gas surrounding
star-forming regions, originating a series of emission lines and a continuum flux. 
The nebular emission is important in the presence of strong starbursts, 
when stellar populations of ages $\tau \le 5-10$ Myr contribute significantly to the
total emission.

For the first time, we implemented this contribution into a semi-analytic code of galaxy 
formation, using the publicly-available models of Molla' et al. (2009). For each age and
metallicity in our grid of input M05 models, we included the nebular emission on top of 
the stellar emission, following the indications of Molla' et al. (2009). 
Notice that this is not stricly physically sensible, in that 
the ionizing photons are not subtracted from the stellar UV spectrum, so that energy
is not conserved for a given stellar population. This however has the advantage of
setting the maximum limit
of the possible contribution of the nebular emission to the total galaxy spectra.

\begin{figure*}
\includegraphics[scale=0.9]{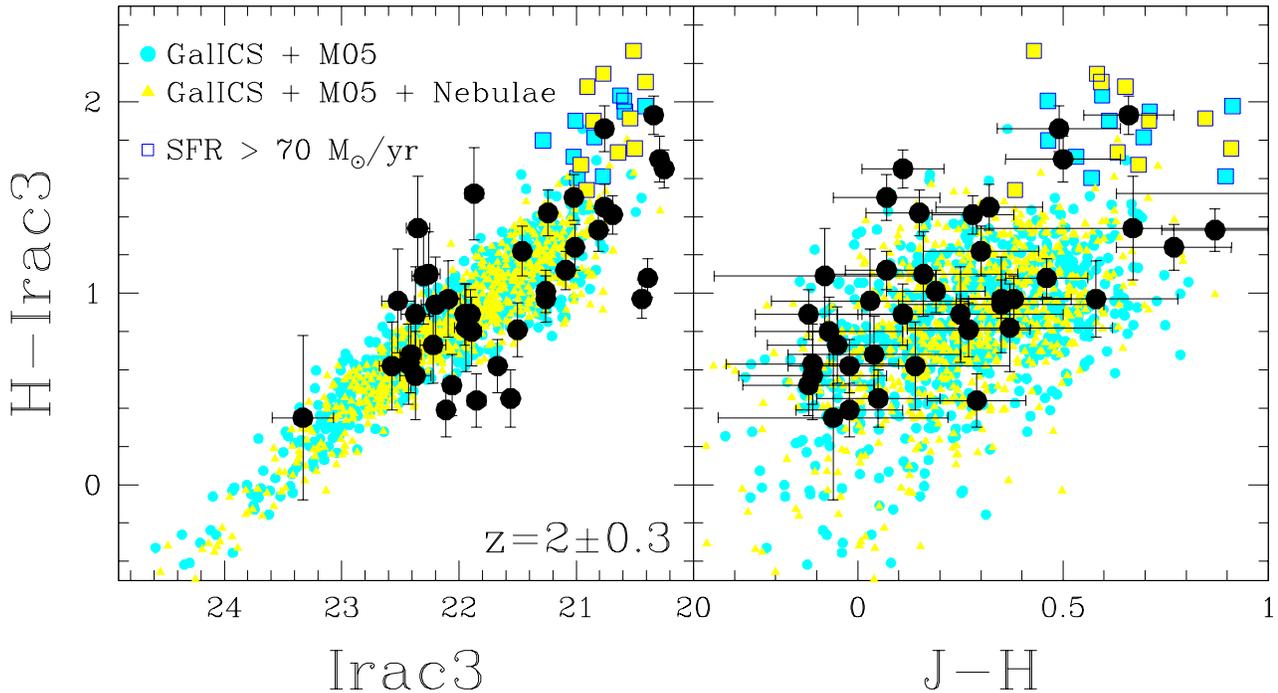}
\caption{The Irac3 $vs$ H-Irac3 colour-magnitude relation (\textit{left panel})
and the  J-H $vs$ H-Irac3 colour-colour relation (\textit{right panel}) at $z=2$ for the run with the
M05 models (\textit{cyan dots})
and for the run with M05 models $+$ nebular emission (as in Molla' et al. 2009; 
\textit{yellow triangles}), compared with
the GOODS-S catalogue (Daddi et al. 2007). The \textit{highlighted squares} represent
galaxies with SFR$>70 \ M_{\odot}/yr$ in both runs. We are only plotting the model galaxies
with reddening, for clarity.}
\label{popstar}
\end{figure*}

The result is presented in Fig.~(\ref{popstar}). The (\textit{left panel}) shows the 
colour-magnitude relation Irac3 $vs$ H-Irac3 and the \textit{right panel}) shows the
colour-colour relation J-H $vs$ H-Irac3 at $z=2$ (observed frame). The M05 run is
represented by \textit{cyan dots}, and the M05 run with nebular emission is 
represented by \textit{yellow triangles}. As before, the model predictions are compared
with the GOODS-S catalogue. Notice that the nebular emission leaves the colours
and luminosities of galaxies virtually unchanged in these runs, except at the very
high-luminosity red end. The \textit{highlighted squares} in the plot represent
galaxies with an instantaneous star-formation rate of SFR$>70 \ M_{\odot}/yr$. For some 
of these objects, the nebular emission increases the H-Irac3 (rest-frame V-K) colour 
by about 0.2-0.4 mags, while it is not so clear for the J-H (rest-frame B-V).

The reason why the nebular emission contribution is not more impressive is
that the semi-analytic model cannot produce very violent starbursts. The high-SFR
tail of the galaxy population at $z=2$ is around 100-200 $M_{\odot}/yr$, 
but these galaxies are relatively rare in the simulation. As stated in 
Leitherer et al. 1999, Zackrisson et al. 2008 and Molla' et al. 2009,
the nebular emission is a main factor in determining the colours of starburst 
galaxies, where very young ($\tau < 5-10$ Myr) stellar populations represent 
a significant fraction of the stellar mass. Hence we conclude that the inclusion
of nebular emission does not affect our results.

\section{Rest-frame K-band luminosity functions}

In the previous Sections we presented a qualitative comparison between the
spectral energy distributions of model and real galaxies, and showed that
the introduction of the TP-AGB in the semi-analytic model substantially improves
the model performance in reproducing the observed galaxies at $z \sim 2$. 

A more quantitative approach is to compute the luminosity function for 
the model galaxy population. After the correct stellar population models are 
implemented in the semi-analytic model and the TP-AGB is included, 
the luminosity function in the near-IR is a good proxy
for the galaxy stellar mass function, therefore representing a meaningful test of 
the mass assembly in the hierarchical model. With one caveat, the role played by AGN feedback.
 
AGN (Active Galactic Nuclei) feedback was introduced in the galaxy formation 
models to turn down the efficiency
of star formation at the high-mass end of the galaxy population, a fundamental recipe 
in order to obtain the correct shape of the luminosity function
(see Benson et al. 2003, Binney 2004, Granato et al. 2004, Silk 2005, Bower et al. 2006, 
Croton et al. 2006). In fact, if stars were
to follow the dark matter halo mass function (which is essentially scale-invariant), the 
models would produce a severe overabundance of massive systems. 
A mechanism is needed in the models to 
suppress star formation in more massive systems, at specific 
points in each galaxy history. 
The most popular solution adopted in the models, a mechanism that can act on a galactic scale 
and heat up the gas, thus preventing star formation, is the energy emission
from the central black hole following gas accretion, the so-called AGN feedback. 
It has the advantage that it is preferentially active in massive objects and at high redshifts
($z > 1-2 $, see for instance Madau et al. 1996, Shaver et al. 1996). 
However, the coupling between the AGN energy release and the gas in galaxies is still 
poorly understood, so that this source of feedback is implemented a-posteriori 
to fine-tune the models, given some set of observational constraints. 

The most widely used calibration data set for semi-analytic models is the luminosity
function at $z=0$, for various photometric bands. Different stellar
population models in the semi-analytic code affect the galaxy spectral energy
distribution and mass-to-light ratio (M05, Tonini et al. 2009),
so before investigating the
predictions at high redshift, a fundamental step is to verify
that the model is still well calibrated at $z=0$. 

\begin{figure}
\includegraphics[scale=0.4]{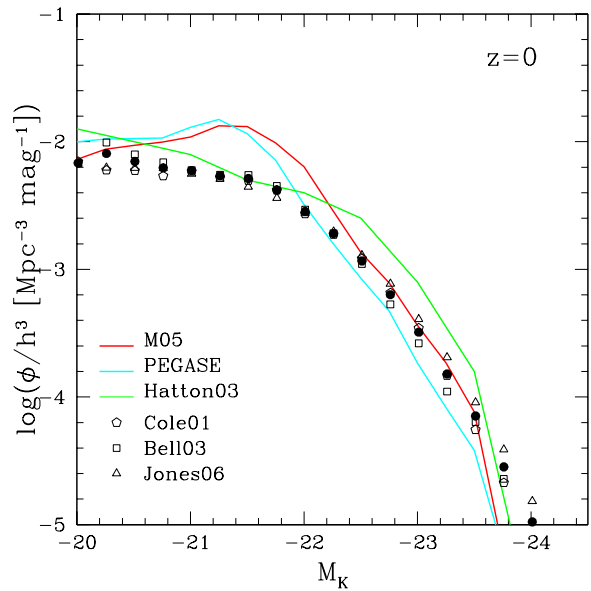}
\caption{The original GalICS rest-frame K-band luminosity function at $z=0$ (Hatton et al. 2003; 
\textit{green line}) compared with data from Cole et al. 2001, Jones et al. 2006, 
Bell et al. 2003  and a combination of these 3 samples (\textit{thick black circles}).
The \textit{thick red line} represents again the M05 run, and the \textit{thick, cyan line} 
represents the PEGASE run.}
\label{LForiginal}
\end{figure}
Fig.~(\ref{LForiginal}) shows the $z=0$ K-band rest-frame luminosity function for 
GalICS as in the original paper (Hatton et al. 2003, \textit{green line}), 
compared to data from Cole et al. (2001; \textit{pentagons}),  
Jones et al. (2006; \textit{triangles}), 
Bell et al. (2003; \textit{squares}) and a combination of these 3 samples 
(\textit{thick black circles}). 
In this magnitude range, the errorbars on each luminosity function are at most as large 
as the spread between the different functions. 
The \textit{red line} is the luminosity function obtained with the M05 run,
and the \textit{thick, cyan line} line represents the PEGASE run.
The difference between the luminosity function in the GalICS $+$ PEGASE run and the
original version of Hatton et al. (2003) is partly due to the difference between the PEGASE and 
Stardust models (used in the original version of the code), and the fact that the current
version of GalICS includes the improved recipe for
AGN feedback introduced by Cattaneo et al. (2006) and the new chemical evolution model
implemented by Pipino et al. (2008). In general, the original version and our two new runs
agree reasonably well with the data, and in particular the TP-AGB allows the model to 
perform better at the high-mass end.  

The LF in the current version of GalICS 
overpredicts the number of galaxies at intermediate luminosities, 
regardless of the stellar population models in use.  
This tension with the data is not critically relevant for the work presented here, 
in that it regards the implementation of physics beyond the SSP models, and it does not affect the 
comparison between the M05 and PEGASE runs. 

\begin{figure}
\includegraphics[scale=0.4]{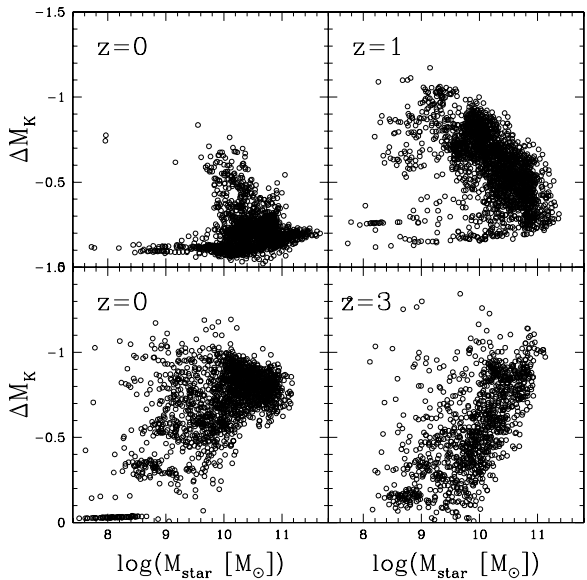}
\caption{The difference between the rest-frame K-band
broadband magnitudes predicted in the M05 and PEGASE runs, $\Delta M_K = M_K (M05) - M_K (PEGASE)$, 
as a function of stellar mass, in 4 redshifts bins from $z=0$ to $z=3$. }
\label{Lmass}
\end{figure}

Notice that, at $z=0$, the K-band luminosity function is 
mildly dependent on the 
input model SSPs. In particular, there is a small difference 
between the M05 run 
and the run with the Stardust models. This 
may originate from the different temperatures of the RGB phase in the two models and also
from residual TP-AGB dominated populations in the simulations at low redshift.
In fact, the amount of intermediate-age population at low redshift is
much lower than at high look-back times; hence
at z=0 the K band mostly traces the Red Giant Branch in galaxies. 
The offset between the M05 and the PEGASE runs is $\sim$ 0.2 mag and is due to 
galaxies that recently had - or are having - star formation.
since for the majority of the galaxies 
the bulk of the stellar populations are old and the TP-AGB phase is subdominant (M05). 

This however does not
remain true at all redshifts, as shown in the previous Sections and in Tonini et al. (2009). The 
TP-AGB becomes the dominant contributor to the near-IR luminosity at $z>1$, so that the 
mass-to-light ratio in the K-band and in the nearing bands is significantly offset between 
the M05 and the PEGASE runs. 
To illustrate this point, Fig.~(\ref{Lmass}) shows the difference between
the predicted K-band magnitudes in the M05 and PEGASE runs, defined as
$\Delta M_K = M_K (M05) - M_K (PEGASE)$, for the same galaxy masses, 
in 4 redshifts bins from $z=0$ to $z=3$. 
At $z=0$ the M05 run produces galaxies on average brighter by 0.3 mag
than the run with the PEGASE recipe,
but at $z>1$ the M05 run gets brighter by 
more than 1 mag, and the offset between the two runs shows a mild dependence on galactic mass.   
This difference is going to be mirrored by the luminosity function at high redshift.

\begin{figure*}
\includegraphics[scale=0.8]{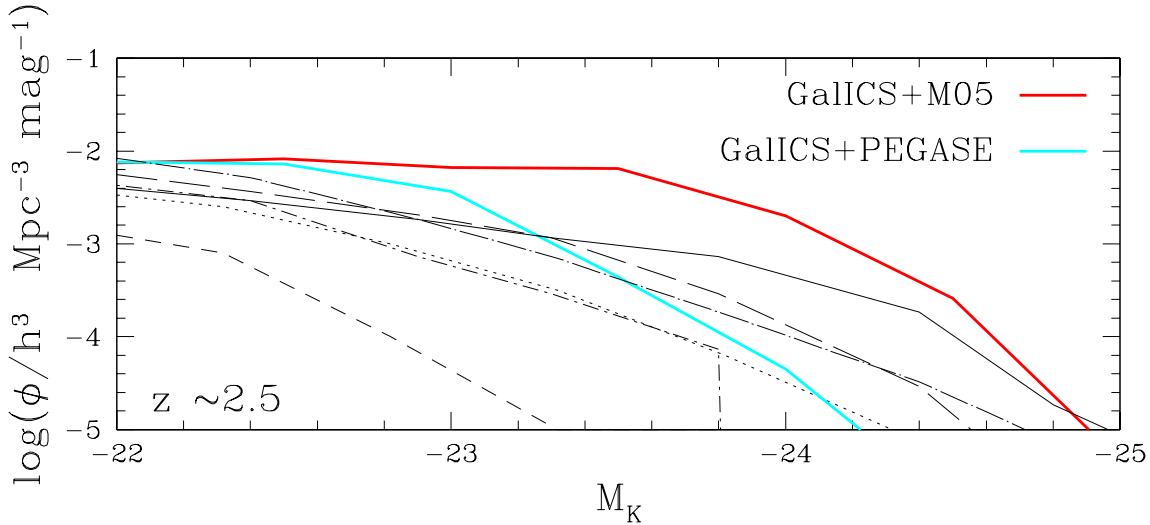}
\caption{The rest-frame K-band luminosity function of the M05 (\textit{thick red line}) and
PEGASE (\textit{thick cyan line}) runs, compared with published LFs in the literature: 
Bower et al. 2006 (\textit{solid}), Cole et al. 2000 (\textit{short-dashed}), 
Baugh et al. 2005 (\textit{dotted}), Menci et al. 2006 (\textit{long dot-dashed}),
Monaco et al. 2007 (\textit{short dot-dashed}) and De Lucia \& Blaizot (2007) 
(\textit{long-dashed}).}
\label{bower}
\end{figure*}

Fig.~(\ref{bower}), shows the $z=2.5$ luminosity functions in the rest-frame K band 
for the M05 and PEGASE runs (\textit{thick red} and \textit{thick cyan} lines respectively). 
As expected, there is an offset of roughly 1 mag between the functions (with the PEGASE run
underpredicting the number of bright galaxies), more pronounced 
at the high-mass end due to the steeper slope of the function and to the mild dependence 
of the offset with galaxy mass. The difference between the two functions is exclusively due
to different stellar population models implemented into the GalICS code. 
The two functions are also compared with the predictions of other semi-analytic models 
in the literature (see the caption for the line-coding), 
all of which make use of either the BC03 models or the GRASIL SSPs of Silva et al. (1998) 
(which implement the TP-AGB, but produce near-IR spectra very similar to BC03; 
P. Monaco, private communication), but differ for the 
various implementations of the baryonic 
physics. In particular, the main factor shaping the bright end of the near-IR luminosity
function is the recipe for AGN feedback. 

Two considerations are important here. The first is that, regardless of the fact that 
all these models are set to match the $z=0$ luminosity function, their predictions 
at high redshift diverge dramatically. This is in part due to the different 
recipes for the baryonic physics adopted in each model. However, semi-analytic models
cannot match the $z=0$ luminosity function without AGN feedback, which in each case 
is implemented ad-hoc to fine-tune 
the model at $z=0$, based on energy arguments at best. The lack of physics in the AGN  
recipe makes it degenerate with other model parameters, expecially at high redshift. 
In fact, AGN activity supposedly 
peaks around $1 < z < 3$, at epochs when the stellar emission is dominated by the TP-AGB in 
the near-IR, so that these two factors compete in shaping the high-redshift luminosity function.  

The second consideration is even more striking. The shift in the luminosity function caused 
by the introduction of the TP-AGB emission in the model is comparable in magnitude to the 
difference introduced by different AGN-feedback recipes. In fact, the M05 and PEGASE runs
actually bracket most of the other semi-analytic models at the high-mass end.
Given that the stellar emission is much better understood and constrained, the importance
of producing realistic and complete stellar population models \textit{before} 
fine-tuning the AGN-feedback recipe is evident. 

\section{Summary and Discussion}

In a recent work (Tonini et al. 2009) we introduced the complete treatment
of the TP-AGB phase of stellar
evolution into a semi-analytic model of galaxy formation, by inserting the 
Maraston (2005) SSP models into the code GalICS. In the work presented here we compared
the predictions on the near-IR luminosities and colours of high-redshift
galaxies with data samples of nearly-passive and star-forming galaxies
around $z \sim 2$. Our main results are: 

$\cdot$ the TP-AGB is fundamental to allow the semi-analytic model to reproduce 
the observed optical and near-IR colours of both nearly-passive and star-forming galaxies
at $z \sim 2$; 
the inclusion of the TP-AGB increases the Irac3 luminosity (rest-frame K)
and shifts the H-Irac3 (rest-frame V-K) colours by more than 1 magnitude;

$\cdot$ without the 
TP-AGB, it is not possible to match the observed galaxy colours and luminosities
by a modification of the dust reddening recipe alone;

$\cdot$ the TP-AGB emission does not alter the optical luminosity and colours
of star-forming galaxies. On the other hand, star formation does not dilute
the TP-AGB emission in the near-IR. Even star-forming galaxies, very blue in 
the optical, can be very red in the near-IR. Therefore the labelling of
red galaxies as 'red and dead' is misleading;

$\cdot$ the nebular emission, produced by young stellar populations, does not
add a significant contribution to the colours of star-forming galaxies, in the 
range of star-formation rates covered by the model; for SFR$>70 \ M_{\odot}/yr$
the rest-frame V-K colour is reddened by 0.2-0.4 mags;

$\cdot$ the predicted mass-luminosity relation is affected by the inclusion
of the TP-AGB; for a given galaxy mass, the rest-frame K-band luminosity
is higher by more than 1 mag at $z>1$. As a consequence, the K-band luminosity 
function predicted by the model with the TP-AGB shifts 
redwards, expecially at the high-mass end, for $z>1$ 
(by $\sim 0.7$ mag at $z \sim 2.5$). The spread 
in the luminosity function between runs with and without the TP-AGB 
is comparable to the scatter caused by different AGN-feedback recipes
in the literature.    

Note that the high-mass end of the luminosity function in the near-IR is dominated
by spheroids, or the progenitors of today's spheroids. If the use of the TP-AGB in the
semi-analytic model shifts the luminosity function by $\sim$1 mag at the high-mass end,
it means that the mass-to-light ratio is lower by a factor of $\sim$2.5 for a given
luminosity. When galaxy masses are inferred from observations by the use of these models,
they are lower by the same factor (as shown in M05). This may rise the question of whether there
is enough mass in spheroids at high redshift to account for the $\sim 50\%$ of stellar mass
in ellipticals measured in the local universe. However, the model correctly predicts the 
stellar mass density at all redshifts, meaning that only the \textit{distribution of galaxy masses} 
is at tension with observations, \textit{if the TP-AGB is not taken into account}. 
In fact, hierarchical models in general predict a faster evolution
of the high-mass end of the stellar mass function than currently inferred from 
observations (see for instance Conselice et al. 2007).
A more accurare derivation of galaxy masses through complete stellar population models
with the TP-AGB, coupled with more accurate predictions from hierarchical models with the 
right input SSP, surely contribute to alleviate the discrepancy.

The inclusion of the TP-AGB allows the semi-analytic model to reproduce the very red end of the 
galaxy population at $z \sim 2$, both for nearly-passive and for star-forming objects. 
It allows the model to do so with a comfortable range of galaxy masses and dust reddening.  
Most importantly, it contributes to a realistic
and comprehensive treatment of the galaxy light emission in galaxy formation models, 
making them a much more precise tool to test our understanding of galaxy assembly.

The implementation of the TP-AGB allows the model to produce, at a given stellar
mass, redder and more luminous galaxies in the near-IR, expecially at high redshift
where the ages of the stellar populations peak around the epoch of maximal emission
from this stellar phase.
In case of nearly-passively evolving galaxies, the model can 
reproduce the red colours and high K-band magnitudes without invoking too large 
stellar masses or too old ages, which would be problematic in the hierarchical context. 
In the case of star-forming galaxies, the TP-AGB still increases
the near-IR luminosity and makes the galaxies redder, without offsetting the 
blue optical colours. Thus, observed red colours in the near-IR do not necessarily imply old ages
and passive evolution, a fact that again would be problematic for the hierarchical
picture at high redshift. In general, the introduction of the TP-AGB in the models
is a step forward in reconciling the hierarchical assembly mechanism with the 
observations of the high-redshift universe. 

\section*{Acknowledgments}
This project is supported by the Marie Curie Excellence Team Grant MEXT-CT-2006-042754
of the Training and Mobility of Researchers programme financed by the European Community. 
We wish to acknowledge Danilo Marchesini, Alvio Renzini, Emanuele Daddi 
and Bruno Henriquez for the 
interesting and useful discussions. We also want to thank Mark Dickinson and Emanuele Daddi 
for letting us use the GOODS-S data sample of star-forming galaxies.
Finally, we wish to thank the Referee for his comments and suggestions, which
helped to improve this work.

\end{document}